\documentclass[preprint,showpacs,preprintnumbers,amsmath,amssymb]{revtex4}
\usepackage{amsmath,amssymb,graphicx}
\usepackage{epstopdf}
\usepackage {amssymb}
\newcommand{\nc}{\newcommand}
\nc{\ba}{\begin{eqnarray}}
\nc{\ea}{\end{eqnarray}}
\newcommand\be{\begin{equation}}
\newcommand\ee{\end{equation}}

\nc{\x}{\vec {\bf  x}}
\nc{\B}{ {\cal  B}}

\input{epsf.sty}

\begin{document}

\title{Inflation from Charged Scalar and Primordial Magnetic Fields? }

\author{ Razieh Emami}
\affiliation{~ Department of Physics, Shahid Beheshti University,
G.C., Evin, Tehran 19839, Iran}

\author{ Hassan Firouzjahi} \email{firouz@ipm.ir}
\affiliation{ School of Physics, Institute for Research in
Fundamental Sciences (IPM), P. O. Box 19395-5531, Tehran, Iran}

\author{M. Sadegh Movahed} \email{m.s.movahed@ipm.ir}
\affiliation{Department of Physics, Shahid Beheshti University,
G.C., Evin, Tehran 19839, Iran, and \\School of Astronomy, Institute
for Research in Fundamental Sciences (IPM),  P. O. Box 19395-5531,
Tehran, Iran}

\begin{abstract}
\vspace{0.5cm}

A model of inflation is presented where the inflaton field is a complex scalar field coupled to a $U(1)$ gauge field. Due to the axial symmetry of the potential, the inflation is driven by the radial direction while the angular field is gauged by $U(1)$. Due to the coupling of the inflaton to the gauge field, a time dependent mass term for the gauge field is generated dynamically and conformal invariance is broken.  We study whether a significant amount of primordial magnetic fields can be generated during inflation by allowing a time-dependent $U(1)$ gauge kinetic coupling.

\vspace{0.3cm}

\end{abstract}

\maketitle
\section{Introduction}
Astronomical observations indicate the existence of magnetic fields
in galaxies and also in cosmological scales with coherence lengths
as big as 1 Mpc and with the magnitude of order $10^{-7}$ Gauss, for
more details see \cite{Grasso:2000wj, Widrow:2002ud,
Giovannini:2002sv}. The origin of the cosmic magnetic fields are not
well-understood. One possible explanation is that they may have
primordial origin which are later amplified by galactic dynamo
mechanism into  the current observed value. The estimation for the
magnitude of the primordial seed  is not certain. Conventionally, it
is assumed that a seed in the range $10^{-25}-10^{-15}$ Gauss at the
time of matter-radiation decoupling may be required for dynamo
mechanism to produce the current observational value. This lower
bound may be relaxed to $10^{-30}$ Gauss in a flat dark energy
dominated  Universe \cite{Davis:1999bt}.

Starting with the work of Turner and Widrow \cite{Turner:1987bw},
during past two decades there were some attempts to obtain
primordial magnetic seed through inflation
\cite{Ratra:1991bn,Dolgov:1993vg,Lemoine:1995vj,Gasperini:1995dh,Calzetta:1997ku,Finelli:2000sh,
Giovannini:2000dj,Giovannini:2000wta,Davis:2000zp,Gasperini:2000tw,Bassett:2000aw,Giovannini:2001nh,Bamba:2003av,Bamba:2008hr,Ashoorioon:2004rs,Ganjali:2005sr,Bamba:2006ga,Giovannini:2007rh, BY, BRM, Salim, Martin, Bamba-Odnitsov, Bamba-G-H, Garretson, MS, Novello, Mota, Massimmo, Kostas, Prokopec, Anupam, Garousi, Mardiz, Bellini, Cea, Campanelli:2008kh,
Demozzi:2009fu}. In particular, inflation can naturally stretch small micro-physical
scales inside a casual patch into the cosmological sizes. This can
provide an elegant mechanism to explain the coherence of magnetic
fields on distances much bigger than astrophysical  scales. The key
obstacle, however, in producing magnetic field during inflation (in
general in a Friedmann-Robertson-Walker (FRW)  background) is the
conformal invariance. To produce magnetic field one has to break the
conformal invariance of the classical electrodynamics. There are
several mechanisms to break conformal invariance. These include:
coupling electrodynamics non-minimally to gravity  which produce a
time-dependent mass for photon \cite{Turner:1987bw, Bassett:2000aw},
introducing a time-dependent gauge kinetic coupling
\cite{Ratra:1991bn, Dolgov:1993vg, Lemoine:1995vj, Gasperini:1995dh,
Giovannini:2001nh, Bamba:2003av, Bamba:2008hr, Ganjali:2005sr,
Bamba:2006ga, Giovannini:2007rh, BY, BRM, Salim, Martin, Bamba-Odnitsov, Bamba-G-H, Demozzi:2009fu} , and  coupling the photon to a
charged scalar field  \cite{Turner:1987bw, Calzetta:1997ku,
Finelli:2000sh, Giovannini:2000dj, Davis:2000zp, Bassett:2000aw}.

In this paper we combine the last two mentioned methods together in
order to break the conformal invariance and obtain an appreciable
amount of seed magnetic field. We consider the model where the
inflaton field is a complex scalar field coupled to a $U(1)$ gauge
field. The conformal invariance is broken dynamically when the
scalar field is displaced from its minimum during inflation.
Furthermore, the time-dependence of the gauge kinetic coupling as in
\cite{Bamba:2006ga, Demozzi:2009fu} can help to produce a
considerable  amount of primordial magnetic fields seed required for
galactic dynamo. As indicated recently in \cite{Demozzi:2009fu}, the
key constraints in models with time-varying gauge coupling are (a): the requirement that the time-dependent gauge
coupling remains small for all time  during inflation, so the
perturbative gauge theory is under control and (b): one must check
that the strong back reaction from gauge field does not destroy the
inflationary background. We shall see that these constraints in
addition with the requirement to have sufficient inflation basically
control the amplitude of the magnetic fields seed produced.

This paper is organized as follows. In section \ref{setup} we present our set up.
In section \ref{magnetic} the magnetic fields produced in our model are studied
followed by discussions and conclusion
in section \ref{conclusion}. While this paper was in its final stage, the
work \cite{Kanno:2009ei} appeared which has some overlaps with our studies.


\section{ Set UP}
\label{setup}

Here we present our set up. It contains a complex scalar field,
$\phi$, coupled to a $U(1)$ gauge field, $A_\mu$, with a time
dependent gauge kinetic coupling, $I(t)$. The action is \ba
\label{action} S= \int d^4 x  \sqrt{-g} \left [ \frac{M_P^2}{2} R -
\frac{1}{2} D_\mu \phi \,  D^\mu  \phi^{\dag} -   \frac{I^2(t)}{4}
F_{\mu \nu} F^{\mu \nu}  - V(\phi, \phi^{\dag}) \right]\, , \ea where
$M_P^{-2} = 8 \pi G$ and $G$ being the Newton constant. The
covariant derivative is given by \ba D_\mu \phi = \partial_\mu  \phi
+ i e \, \phi  \, A_\mu \, , \ea where $e$ is the dimensionless gauge
coupling of $A_\mu$ to $\phi$. As usual, the gauge field strength is
given by
\begin{eqnarray} F_{\mu \nu}=
\partial_\mu A_\nu -
\partial_\nu A_\mu \, . \end{eqnarray}
Motivated by
\cite{Demozzi:2009fu,Bamba:2006ga} we have entertained the
possibility that the gauge kinetic coupling is time dependent. This
may arise for example in string theory setup where the gauge kinetic
action is coupled to the dilaton which is running with time. As
suggested in  \cite{Demozzi:2009fu} one may also consider that the
time dependence of $I(t)$ originates from the time dependence of
$e(t)$. This may be obtained by replacing $A_\mu \rightarrow
A_\mu/e(t)$. But this has the problem that the kinetic action
$F_{\mu \nu}F^{\mu \nu}$ becomes modified under  $A_\mu \rightarrow
A_\mu/e(t)$ and the gauge  invariant is lost. At the
phenomenological level, we shall take $e$ to be constant and assume
the time dependence of $I(t)$ originates from the coupling of the
gauge kinetic action to other fields such as dilaton. However,  the
dynamics of the dilaton should be such that it does not destroy the
slow-roll properties of the inflationary potential. This may be a
non-trivial task, but we shall take the phenomenological view of
\cite{Demozzi:2009fu, Bamba:2006ga} and proceed with action
represented in Eq.  (\ref{action}).

We work with potentials which have axial symmetry where $V$ is only
a function of $\phi \phi^{\dag}=  |\phi |^2$. It is more instructive
to decompose the inflaton field into the radial and angular parts
\ba \phi(x) = \rho(x) \,  e^{i \theta(x)}\, ,  \ea so $V=V(\rho)$. As
usual, the action, Eq.  (\ref{action}), is invariant under local
gauge transformation \begin{eqnarray} \label{transformation} A_\mu
&\rightarrow& A_\mu - \frac{1}{e} \partial_\mu \epsilon(x)\nonumber\\
\theta &\rightarrow& \theta + \epsilon(x) \, .
\end{eqnarray} With this decomposition, Eq.  (\ref{action}) is
transformed into: \ba \label{action2} S= \int d^4 x \sqrt{-g} \left
[ \frac{M_P^2}{2} R - \frac{1}{2}
\partial_\mu \rho
\partial^\mu \rho-
\frac{\rho^2}{2}  \left( \partial_\mu \theta + e A_\mu  \right)
\left( \partial^\mu \theta + e A^\mu  \right) - \frac{I(t)^2}{4}
F_{\mu \nu} F^{\mu \nu}  - V(\rho) \right] \ea The corresponding
Klein-Gordon equations of motion are: \begin{eqnarray}
\label{theta-Eq} 0&=&\partial_\mu\,  J^\mu
\\
\label{rho-Eq} 0&=&\partial_\mu \left[  \sqrt{-g} \partial^\mu \rho
\right] - \frac{ J_\mu J^\mu}{\rho^3 \sqrt{-g} }   -\sqrt{-g}\,
V_\rho \, \, ,
\end{eqnarray}
accompanied by with the Maxwell's equation
\ba
\label{Maxwell}
\partial_\mu \left(  \sqrt{-g}\, I(t)^2\, F^{\mu \nu} \right) = e J^\nu \,
\ea where the current, $J^\nu$, is defined by \ba J^\nu \equiv
\rho^2 \sqrt{-g} \left( \partial^\nu \theta + e A^\nu  \right)  \, .
\ea The conservation of $J^\mu$ from Eq.  (\ref{theta-Eq}) is a
manifestation of the axial symmetry imposed on $V$. Interestingly,
Eq.  (\ref{theta-Eq}) is not independent from Maxwell's equation,
where $F^{\mu \nu}$ being anti-symmetic leads to $\partial_\mu
\partial_\nu F^{\mu \nu} = \partial_{\mu} J^{\mu}=0$.

Finally, the stress energy momentum tensor, $T_{\alpha \beta}$, for
the Einstein equation, $G_{\alpha \beta} = 8 \pi G \, T_{\alpha
\beta}$, is: \ba \label{energy-momentum} T_{\alpha \beta} =
\frac{-I(t)^2}{4} g_{\alpha \beta} F_{\mu \nu} F^{\mu \nu} +I(t)^2
F_{\alpha \mu} F_\beta^{\, \mu} +  \partial_\alpha \rho
\partial_\beta \rho +  \frac{J_\alpha J_\beta}{\rho^2 | g|} -
g_{\alpha \beta} \left[ \frac{1}{2} \partial_\mu \rho
\partial^\mu \rho +  \frac{J_\mu J^\mu}{2 \rho^2 | g|}  +V
\right]\, .
\ea

At the background level, we start with the isotropic and homogeneous
FRW space-time with the metric \ba \label{metric} ds^2 = - dt^2 +
a(t)^2 d \x^2 \, .  \ea Considering the $\nu=0$ component of the
Maxwell equation (\ref{Maxwell}) implies that \ba J^0 = \rho^2 a^3
(\dot \theta + e A^0) =0 \,  . \ea This indicates that the total
electric charge  is zero \cite{Dolgov:1993mu}. This is somewhat
similar to usual angular momentum conservation where in the absence
of gauge field, an axial symmetric potential leads to angular
momentum conservation $\partial_t (a^3 \rho^2 \dot \theta)=0$.

The gauge field $A_\mu$ is invariant under gauge transformation Eq. (\ref{transformation}). We need to fix the gauge to study the
physical independent degrees of freedom. We choose the
Coloumb-radiation gauge where $A_0 = \partial^i A_i=0$. With
$A_0=0$, from $J^0=0$, one obtains that $\dot \theta =0$ at the
level of background. The $\nu=0$ component of the perturbed Maxwell
equation (Eq. (\ref{Maxwell})) with the above gauge implies that
$\delta \theta =0$. This is also consistent with the perturbed
angular field equation (Eq. (\ref{theta-Eq})).

A non-zero spatial component of vector field at the background level
may produce large anisotropies.
In \cite{Kanno:2009ei} the mechanism of magnetic seed production in the presence of an anisotropic gauge field was studied to some extent. It was shown that an anisotropic gauge field background will modify the inflation dynamics. It is found that the final produced magnetic field  becomes highly suppressed as compared to the isotropic background. Also it is argued that a large anisotropic background may produce large anisotropies on CMB which may be detectable and need further studies. Here, in order to simplify the analysis and,
in the light of \cite{Kanno:2009ei}, to have an optimum  primordial magnetic field  production, we assume that $\vec A=0$ at the background
level. We need to check if this is a consistent solution of the
background inflationary dynamics. To this end,  we note that in our
isotropic background, $F_{ij}= J^i=0$ and
one can easily check that $\vec A= \dot {\vec A}=0$ is a consistent solution of
Eqs. (\ref{theta-Eq})-(\ref{Maxwell}) and the Einstein equations. In other words, the
$\rho$ field and the gauge field equations of motion decouple from each other. One also has to check that this ansatz is stable against perturbations. To see this, we note that the mass squared term created for the gauge field via
spontaneous symmetry breaking (the Higgs mechanism) is positive, given by $e^2 \rho^2 /2$,
which indicates there is no tachyon in the mass spectra of the gauge field.

With this ansatz, at the background level, the independent equations of
motion are \begin{eqnarray} \label{background} 0&=&\ddot \rho + 3 H
\dot \rho +
V_\rho\nonumber\\
 3 H^2 M_P^2& =& \frac{1}{2} \dot \rho^2 + V(\rho)
\, .  \end{eqnarray}The interesting result is that the inflation is
completely driven by the radial field $\rho$ and the angular field
$\theta$ is gauged by $U(1)$. As in conventional single field models
of inflation, one can choose $V(\rho)$ to be flat enough to support
an extended period of inflation. To be specific, below we consider
the chaotic inflationary models. For potential $ V= m^2 \rho^2 /2$,
the number of e-folding, $N_e$, is given by $4 \, M_P^2 \, N_e
\simeq  \rho_i^2 $, where $\rho_i$ is the initial value of the
scalar field. To solve the flatness and horizon problem, we may take
$N_e =60$, corresponding to $\rho_i \sim 10 M_P$. Furthermore, to
fit the WMAP normalization for the density perturbation, $P_{\cal R}
\sim 2.4 \times 10^{-9} $, one requires that $m \sim  6 \times
10^{-6} M_P$. For the quartic inflationary potential $V=
\lambda\phi^4/4$,
to fit the the data one requires that $\lambda \simeq 10^{-13}$ and $\rho_i \sim 10 M_P$.


\section{Magnetic Field Production During Inflation}
\label{magnetic}

Here we study magnetic field production during inflation. As is
well-known, magnetic field production in FRW backgrounds is heavily
suppressed due to the conformal invariance. To produce a significant
amount of magnetic field, conformal invariance of the Maxwell
equation should be broken. As we see from Eqs. (\ref{action2}) and
(\ref{Maxwell}), the conformal invariance is broken via spontaneous
breaking of $U(1)$ symmetry  where a mass term for the photon is
generated dynamically through the coupling of $\rho$ field to
$A_\mu$. The possibility of generating magnetic field during
inflation due to $U(1)$ symmetry breaking by a complex field is
studied extensively in literature \cite{Turner:1987bw,
Finelli:2000sh, Davis:2000zp, Bassett:2000aw, Giovannini:2001nh}. In
these models  the complex scalar field is usually different than the
inflaton field  and it is assumed that the mechanism of magnetic
field production happens on top of the   inflationary background
without affecting  the dynamics of inflation. However, in our model
we assume  that the inflaton field is the same as the complex scalar
field, i.e. the field $\rho$. This has the advantage that the
parameters controlling the magnitude of magnetic fields production
are directly related to the parameters that control the inflationary
predictions, such as $N_e$ and density perturbations.

The vector perturbations are generated only quantum mechanically
during inflation. To study their evolution and production, we go to
Fourier space where the equation for the spatial components of the
vector field is \ba \label{Ak-Eq}
 A_{i\, k}'' + 2 \frac{I'}{I} A'_{i\, k} +
  \left( k^2  + \frac{e^2}{I^2}\,  a^2 \rho^2 \right ) A_{i\, k} =0   \, ,
\ea where $k$ is the comoving wave number. Hereafter, the prime
indicates the derivative with respect to the conformal time $d \tau
= dt/a(t)$. This equation represents a damped harmonic oscillator
with a time-dependent mass. One can get rid of the first derivative
term in Eq. (\ref{Ak-Eq}) using $\vec v_k = I(\tau) \vec A_k$, so
\ba \label{veq1} v_{ k}'' + \left( k^2 - \frac{I''}{I} +
\frac{e^2}{I^2}\,  a^2\, \rho^2  \right) v_{ k}=0 \, , \ea For
convenience we omits the indices $i$. As in \cite{Demozzi:2009fu} we consider
the following ansatz for the gauge kinetic coupling
 \ba
\label{Ieq} I(\tau) = I_f\, \left(\frac{a}{a_f} \right)^{-p} \, ,\ea
where $I_f$ and $a_f$ are the values of $I(\tau)$ and the scale
factor at the end of inflation, respectively.  We expect that $I_f
\lesssim 1$. Noting that the gauge kinetic coupling is inversely
related to $I$, a negative value of $p$ corresponds to the case when
gauge coupling is very large at the beginning of inflation and
perturbative analysis is not reliable in the beginning of inflation.
On the other hand, for a positive value of $p$ the gauge coupling is
very small at the beginning of inflation and with the assumption
$I_f \lesssim 1$ the perturbative gauge theory is under control for
all time. For this purpose, we take $p \geq 0$ in our analysis
below. Plugging the ansatz Eq. (\ref{Ieq}) into Eq. (\ref{veq1})
yields \ba \label{veq2} v_k'' + \left( k^2 - \frac{p (p-1)}{\tau^2}
+ \frac{\beta}{\tau^{2p+2}}  \right) v_k=0 \, , \ea where \ba \beta
\equiv \frac{e^2 \langle \rho^2 \rangle}{I_f^2  a_f^{2p}
H^{2p+2} } \, .\ea To obtain Eq. (\ref{veq2}) the relation $a \simeq
- 1/ H \tau$ is used during the slow-roll inflation, where $H$ is
the Hubble constant during inflation. Also we assume that $\rho$ is
changing very slowly during inflation  so we replace $\rho^2$ by its
average value $\langle \rho^2 \rangle$.

The amplitude of the magnetic field , $\delta_B$, at the end of
inflation is given by  \cite{Demozzi:2009fu} \ba \label{deltaB}
\delta_B(\tau_f) = \frac{k^{5/2} |v_{k}(\tau_f)| }{ \sqrt2 \pi a_f^2
I_f} \, ,  \ea where $|v_{k} (\tau_f) |$ is the magnitude of
$v_k$ at the end of inflation.

Eq. (\ref{veq2}) can not be analytically solved for arbitrary value
of $p$. Before considering general positive value of $p$, first we
consider the particular case of $p=0$ corresponding to a constant
gauge kinetic coupling, $I=1$.

\subsection{Constant Gauge Kinetic Coupling, $p=0$}
For $p=0$ the solution of Eq. (\ref{veq2}) are given in terms of
Hankel functions \begin{eqnarray} v_k&=&A_k\nonumber\\&=&
\frac{\sqrt{\pi | \tau|}}{2} e^{i \pi( 1+ 2 \nu_0 )/4} \left [ b_1
H_{\nu_0}^{(1)} ( k | \tau| ) + b_2 H_{\nu_0}^{(2)} ( k | \tau| )
\right] \, , \label{hankle1}  \end{eqnarray} where $\nu_0^2 = 1/4 -
\beta$. To match the initial vacuum state \ba \label{vac} v_k  =
\frac{e^{-i k \tau}}{\sqrt{2k}} \, , \ea one requires that $b_1=1$
and $b_2=0$ in Eq. (\ref{hankle1}). On the other hand, for long
wavelength modes, $k | \tau| \rightarrow 0$, and by using the
asymptotic relation \ba \label{asym}
 H_{\nu_0}^{(1)} ( k | \tau| ) \rightarrow   - \frac{i}{\pi} \Gamma(\nu_0)
\left( \frac{k |\tau|}{2} \right)^{-\nu_0} \, , \ea at the end of
inflation one obtains \ba \label{n=0} \delta_B(\tau_f) \simeq
\frac{\Gamma(\nu_0)\,  2^{\nu_0-3/2}}{\pi^{3/2} I_f} \, H^2
\left( \frac{a_f  H}{k} \right)^{\nu_0 - 5/2} \,  , \ea where
$\Gamma(x)$ is the Gamma function.  Noting that the real value of
$\nu_0$ can not exceed $1/4$ the amplitude of magnetic field
in Eq. (\ref{n=0}) is hugely scale suppressed.


To calculate the amplitude of magnetic fields after inflation we
should take into account the expansion of the Universe which dilutes
the magnetic field  energy density, $B_i B^i$, like radiation and
$\delta_B$ decreases like $a^{-2}$. Assuming that the preheating and
reheating happens instantly, followed by a radiation-dominated
Universe, one obtains that  $a_{dec}/a_f \simeq \sqrt{H M_P}/T_{dec}$
where $T_{dec}$ and $a_{dec}$ are the values of the  temperature and
scale factor at the time of decoupling, $t_{dec}$, respectively. For
$H \simeq 10^{-6} M_P$ one obtains  $a_f \sim 10^{-29}$ and
$a_{dec}/a_f \simeq 10^{26}$ where we have set the magnitude of the
scale factor today equal to unity, $a_0=1$. Putting all together,
with $H^2 \simeq 10^{-12} M_P^2 \sim 10^{46}$ Gauss, the amplitude of the
magnetic fields at the time of decoupling is  $\delta_B(t_{dec})
\simeq 10^{-6} \times 10^{11(2 \nu_0-5)} \lesssim 10^{-50}$ Gauss on
comoving scales $k^{-1}= 1 Mpc\sim 10^{38} GeV^{-1}$. This seed
magnetic field is  too small to be  amplified by galactic dynamo
mechanism into the current observed values.

One may compare  the case $p=0$ here to the models studied
in \cite{Turner:1987bw, Bassett:2000aw, Demozzi:2009fu} where the
gauge field is coupled to the Ricci scalar via  the mass term $\beta
R A_\mu A^\mu/2$ and the gauge invariance and the conformal
invariance are broken explicitly. However, in our model the
gauge invariance and the conformal invariance are broken only
dynamically (spontaneously) when the inflaton field is dislocated
from the minimum during inflation. In their cases the conformal
coupling parameter $\beta$ can have a negative value so a negative
mass squared can be generated for the gauge field. This results  in
an amplification of the magnetic field, and one has to check that
its strong back reactions do not destroy the background inflationary
setup \cite{Demozzi:2009fu}. However, in our case $\beta>0$ and a
positive mass squared term is created for the gauge field and the
magnetic field is more suppressed compared to the case of a massless
gauge field where $\beta=0$ and $\nu=1/2$.

\subsection{Magnetic Fields from Time Varying Gauge Kinetic Coupling,
$p>0$}

To keep the gauge kinetic coupling under perturbative control, we
take $p>0$. As it is obvious, Eq. (\ref{veq2}) can not be solved
analytically, so we rely on the numerical solution and asymptotic
behaviors.

During early stage of inflation the first two terms inside the
bracket of Eq. (\ref{veq2}) dominate. The last term dominates
only at the final stage of inflation when $\tau \rightarrow 0$.
For the last term to dominate over the second term one requires that
 \ba \label{Omega}
\Omega\equiv \beta \, |\tau_f|^{-2p} = \frac{e^2
\langle\rho^2\rangle}{I_f^2 H^2} \gg1\, .\ea To have an estimate
of $\Omega$, consider the inflationary potential $V=m^2 \rho^2/2$.
Using the relation $6H^2 M_P^2 = m^2 \rho^2 $, one obtains \ba
\Omega = 6 \left( \frac{e M_P}{m I_f} \right)^2 \, . \ea To fit
the COBE normalization of density perturbation, $m\sim 10^{-6}M_P$
so, for $e \sim I_f \sim1$, $\Omega$ can be as large as
$10^{12}$. Increasing(decreasing) the ratio $e/I(\tau_f)$ the magnitude of
$\Omega$ increases(decreases). On the other hand, for
$\lambda \phi^4/4$ inflationary model, one obtains \ba \Omega =
\frac{12}{\lambda} \left(  \frac{e M_P}{I_f\, \rho}
 \right)^2 \, .
\ea For $\lambda \simeq 10^{-13}$ and $\rho \simeq 10M_P$, $\Omega $
can be as large as $10^{12}$ like in the previous example.

In the early stage of inflation when the term containing $\beta$ is
negligible in Eq. (\ref{veq2}), the asymptotic solution is \ba
\label{short} v_k^{\rm{early}}(\tau) = \frac{\sqrt{\pi | \tau|}}{2} e^{i (
1+ 2 \nu ) \pi/4} H_{\nu}^{(1)} ( k | \tau| )  \,  \ea where
$\nu \equiv p-1/2$. At the final stage of inflation, the term
containing $\beta$ dominates and the asymptotic solution of Eq.
(\ref{veq2}) is given by: \ba \label{long} v_k^{\rm{final}}(\tau) =
\frac{i \sqrt{\pi | \tau|}}{2} \left[ c_1 \, H_\mu^{(1)} \left(
\frac{\sqrt \beta | \tau|^{-p}}{p} \right )+ c_2 \, H_\mu^{(2)}
\left( \frac{\sqrt \beta | \tau|^{-p}}{p} \right ) \right] \, , \ea
where $\mu = 1/2p$ and $c_1$ and $c_2$ are the Bogoliubov
coefficients. One needs to match solutions $v_k^{\rm{early}}$ and
$v_k^{\rm{final}}$ at the transition time, $\tau_c$, where \ba
\label{tauc}
  \beta \, |\tau_c|^{-2 p}  = | p(p-1)|\, . \label{tau}
\ea Since $\tau_c$ is expected to be close to the end of inflation,
then $k \tau_c \rightarrow 0$ and $\tau_c$ given above is
independent of $k$.

Demanding that both $v_k$ and $v_k'$ to be continuous at $\tau_c$ fixes the coefficients
$c_1$ and $c_2$
\ba
c_{1,2} = \frac{\pm \pi}{4 p} e^{i ( 1+ 2 \nu ) \pi/4} \left[  \sqrt{\beta} | \tau_c|^{-p} H_\nu^{(1)} \,
H_\mu^{(2,1)'} + k | \tau_c| \,   H_\nu^{(1)'} \,  H_\mu^{(2,1)  } \right] \, ,
\ea
where the arguments of $H_\mu$ are $\sqrt \beta |\tau_c|^{-p}/p$ while that of $H_\nu$
are $|k \tau_c|$.

 It is instructive to introduce the dimensionless variable $x \equiv k |\tau|$,
and rewrite Eq. (\ref{tau}) as
\begin{equation}
\label{xc-Eq}
x_c=\left[\frac{\Omega}{|p(p-1)|}\right]^{1/2p}x_f \, ,
\end{equation}
where $x_f=k|\tau_f|$.

\begin{figure}
\vspace{-1cm}
\includegraphics[height=10cm, width=12cm]{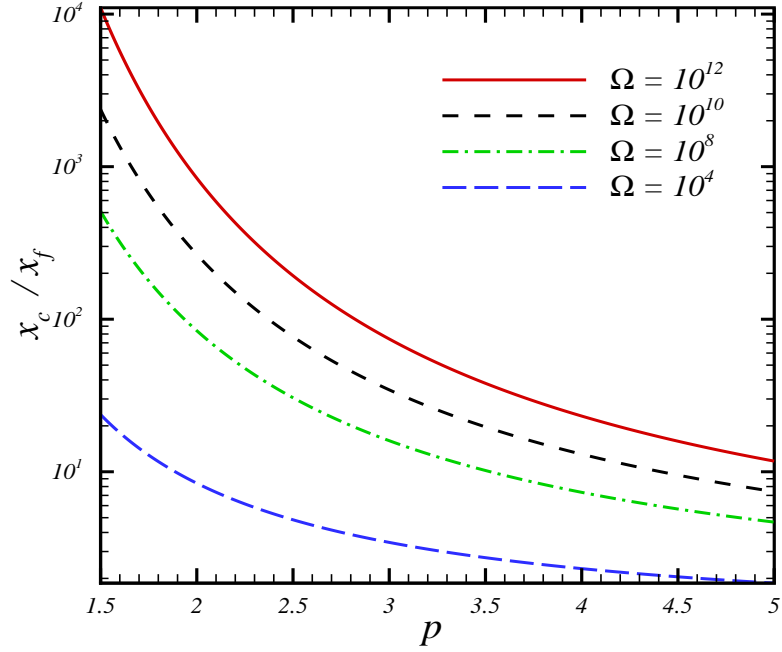}
\caption{ Here we plot the behavior of $x_c\equiv k |\tau_c|$ versus $p$ given by  Eq. (\ref{xc-Eq})
for various values of $\Omega$. We took $x_f=10^{-22}$.} \label{fig4}
\vspace{0.5cm}
\end{figure}

Figure \ref{fig4} shows the value of $x_c/x_f$ as a function of
$p$. For a given value of $p$, by increasing $\Omega$ the value of
$x_c$ goes to the larger values as expected. We note that  with $k^{-1}= 1 Mpc$ and
$H \simeq 10^{-6} M_P$ one has $x_f \simeq 10^{-22}$.
This indicates  that $x_c \rightarrow 0$ as verified from Eq. (\ref{tau}) and also from Figure 1. Physically, this means that the last term in Eq. (\ref{veq2}) dominates at the late stages of inflation.

\begin{figure}
\vspace{-1cm}
\includegraphics[height=10cm, width=12cm]{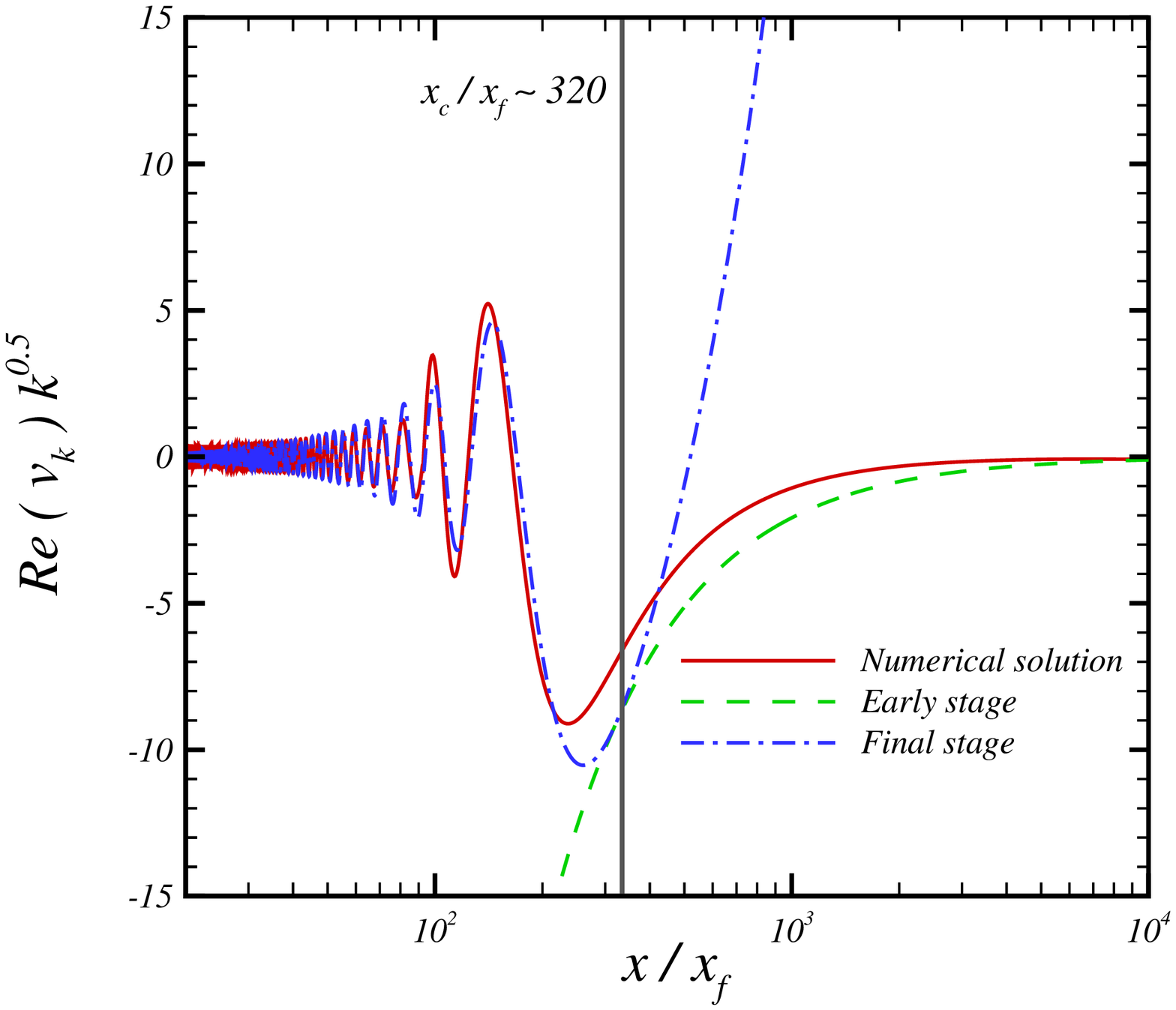}
\caption{The numerical solution of the real part of  Eq. (\ref{veq2}) as a function of $x = k | \tau|$. To make the presentation more transparent,  the vertical axis is scaled down by a factor of $10^{-25}$.  The solid (red) curve indicates the full numerical solution whereas the other two curves indicate the early and the late time solutions, given by
Eqs. (\ref{short}) and (\ref{long}).
 We took $p=2.3$,  $\Omega=10^{12}$ and  $x_f=10^{-22}$. The transition point, $x_c$,
  calculated from Eq. (\ref{tau}), is indicated by the vertical line.
  We see that at $x_c$ the solutions  (\ref{short}) and (\ref{long}) match smoothly and are in a good agreement with the full numerical solution.}
  \label{fig2}\vspace{0.5cm}
\end{figure}

The full Numerical solution of Eq. (\ref{veq2}) for typical value of $p=2.3$
and $\Omega =10^{12}$ along with the asymptotic behaviors at the initial stage, Eq. (\ref{short}), and the asymptotic solution at the final stage, Eq. (\ref{long}),  are
plotted in Figure \ref{fig2}. One can check that the approximate solution Eq. (\ref{long})
is in a good agreement with the full numerical solution. The transition point, $x_c$,
where the two solutions Eq. (\ref{long}) and (\ref{short}) are matched smoothly
together can be seen in Figure 2.

As explained above, we expect that $k \tau_c \rightarrow 0$, which
can be used to simplify the value of $c_1$ and $c_2$. Using the
asymptotic form Eq. (\ref{asym}), one obtains that \ba \label{c12} |
c_1 | \,  \sim | c_2 | \sim \,  | k\,  \tau_c|^{-\nu} \, , \ea and \ba
\label{vf1} |v_k^{\rm{final}}(\tau_f) | \sim \sqrt{|\tau_f|} \,
{\Omega}^{-\frac{1}{4}} | k\,  \tau_c|^{-\nu} \, , \ea where to get
the final answer use was made of the large argument limit of the
Hankel functions $ H_\mu(x) \sim x^{-1/2} $ for $x \gg 1$. One the
other hand, using Eq. (\ref{tauc}) one can express $\tau_c$ in favor
of $\tau_f$ via $\tau_c/\tau_f \simeq \Omega^{1/2 p}$. Putting all
together, and noting that $\nu= p-1/2$ and neglecting factors of
order unity one has \ba \label{vf2} |v_k^{\rm {final}}(\tau_f) |
\sim \sqrt{|\tau_f|} \,   \Omega^{-\frac{3}{4}+\frac{1}{4p}}  |
k \tau_f|^{-\nu} \, , \ea which from Eq. (\ref{deltaB}) results in
\ba \label{deltaB2} \delta_B(\tau_f) \sim
\Omega^{-\frac{3}{4}+\frac{1}{4p}} H^2 \left(  \frac{a_f H}{k}
\right)^{p-3} \, . \ea

As expected, up to term containing the factor of $\Omega$, this is
the same as in \cite{Demozzi:2009fu}. Since
$\Omega \gg 1$, one may expect that our value of
$\delta_B(\tau_{dec})$ would be more suppressed compared to the
result of \cite{Demozzi:2009fu}. On the other hand, due to the
presence of $\Omega$, the allowed range of $p$ increases as compared
to that in \cite{Demozzi:2009fu}.  One has to check whether the
increase in range of $p$ can compensate the suppression from large
$\Omega $ effect. In the examples below we will show that
unfortunately Eq. (\ref{deltaB2}) gives smaller value for the
magnetic fields than those obtained in \cite{Demozzi:2009fu} in the
absence of $\Omega$.

As in  \cite{Demozzi:2009fu}, there is an upper bound on $p$,
determined by the requirement that the energy density from the
vector field, $\epsilon_{EM}$, does not exceed the background
inflationary energy density. This is equivalent to $\epsilon_{EM} <
M_P^2\, H^2$. There are two sources for $\epsilon_{EM}$ coming from
Eq. (\ref{energy-momentum}). The first one, which was also
considered in \cite{Demozzi:2009fu}, is $ \epsilon^{(F)}_{EM}\equiv
-\frac{1}{4}I(t)^2 F_{\alpha \beta} F^{\alpha \beta} + I(t)^2 F_{0
\alpha} F_0 ^ {\, \alpha}$ and the second one is the contribution
from the gauge coupling $e$ in the form of   $\epsilon^{(e)}_{EM}
\equiv J_{\alpha} J^{\alpha}/\rho^2 |g| =  e^2 \rho^2 A_i A^i/2$.
Below we show that $\epsilon^{(e)}_{EM}$ dominates over
$\epsilon^{(F)}_{EM}$ by a factor of $\Omega$. Following
\cite{Demozzi:2009fu}, one obtains \ba
\epsilon^{(e)}_{EM}(\tau_f) &=& \frac{1}{2}e^2 \rho^2 \langle 0| A_i A^i|0\rangle  \nonumber\\&\sim& \frac{\Omega H^2}{4 \pi^2 a_f^2} \int_{a_i H}^{a_f H} dk\, k^2 |
v_k(\tau_f)|^2  \, , \ea where $a_i$ is the initial size of the
scale factor at the start of inflation. Using Eq. (\ref{vf1}), one
obtains \ba \label{eEM} \epsilon^{(e)}_{EM}(\tau_f)  \sim
\Omega^{-\frac{1}{2}+\frac{1}{2p}} \,  \frac{H^4}{4 \pi^2} \times
\left\{
\begin{array}{c}
\frac{1}{2-p}    ~~~~~~~~~~~~~ \quad \quad p<2 \\
\ln \left( \frac{a_f}{a_i}  \right)  ~~~~~~~~~ \quad \quad p=2  \\
\frac{1}{p-2}   \left( \frac{a_f}{a_i}  \right) ^{2 (p-2)}  \quad \quad p>2
\end{array}
\right. \ea
where to get the final result, the relation $\tau_f
\simeq -1/a_f H$ is used during slow-roll inflation. On the other
hand, $\epsilon^{(F)}_{EM} $ is given by \cite{Demozzi:2009fu}
$\epsilon^{(F)}_{EM}  \sim a_f^{-4}  \int_{a_i H}^{a_f H} dk \,  k^2
\,  | v_k'(\tau_f)|^2  $. One can show that $\epsilon^{(F)}_{EM}
\simeq \Omega^{-1} \epsilon^{(e)}_{EM}$  and with $\Omega \gg1$, the
requirement  $\epsilon^{(e)}_{EM} < M_P^2\, H^2$ is the stronger
constraint that has to be satisfied.

From Eq. (\ref{deltaB2}), to have a large enough magnetic field, it
is optimum to have $p > 2$. However, the magnitude of $p$ is
controlled by the value of $\Omega$ and $a_f/a_i$. For example, with
$H \sim 10^{-6} M_P$, $N_e=\ln(a_f/a_i)= 60$ and $\Omega=10^{12}$
which is naturally obtained in chaotic inflationary models, the
constraint $\epsilon^{(e)}_{EM} < M_P^2 H^2$ results in $p \lesssim
2.3$.  This results in $\delta_B(t_{dec}) \sim 10^{-29}$ Gauss on
$k^{-1} \sim$ Mpc scales. This value of magnetic seed may be
marginally acceptable in the light of \cite{Davis:1999bt}. Reducing
$\Omega$ results in an increase in $\delta_B(t_{dec}) $. For
example, choosing $\Omega=10^{4}$, one obtains $\delta_B(t_{dec})
\sim 10^{-25}$ Gauss. With $I_f \sim 1$, this value of $\Omega$ corresponds to
$e \sim 10^{-4}$ which indicates a severe fine-tuning. On
the other hand, in the absence of $\Omega$ factor, one obtains $p
\lesssim 2.2$ and  $\delta_B(t_{dec}) \sim 10^{-23}$ Gauss.


\section{conclusion}
\label{conclusion}

We considered a model where  the inflaton field is a charged scalar
field coupled to a $U(1)$ gauge field  with a time-varying gauge
kinetic coupling. The gauge field is frozen in the classical
background and is excited quantum mechanically. The amplitudes of
magnetic fields produced down to the end of inflation and at the
time of decoupling are calculated. Our model has features in common
with models such as in \cite{Turner:1987bw, Finelli:2000sh,
Davis:2000zp, Bassett:2000aw, Giovannini:2001nh} where a charged
scalar field is present in an inflationary  background. However,
since in our model the charged scalar field itself is the inflaton
field, the prediction of magnetic fields are directly controlled by
the inflationary parameters such as $e$, $N_e$ and $m$ in $m^2
\rho^2$ or $\lambda$ in $\lambda \phi^4$ models.  As in
\cite{Demozzi:2009fu}, the main constraint controlling the magnitude
of $\delta_B$ comes from the requirement that the back reaction from
the gauge field does not destroy the inflationary background. This
in turn imposes an upper bound on the range of parameter $p$. Our
model predicts values of $\delta_B(t_{dec})$ somewhat smaller than
what is obtained in \cite{Demozzi:2009fu} in the absence of charged
scalar field. However, our model predicts  a higher value of
spectral index parameter, $p$, for the primordial magnetic field.
With natural parameter values, our model predicts that magnetic
field at the order of $\delta_B(t_{dec}) \sim 10^{-29}$ Gauss can be
created on $k^{-1} \sim$  Mpc scales. This value may be marginally
acceptable in the case of very efficient  dynamo mechanism.

In this model important issues such as the amplification of magnetic
fields during preheating and subsequent suppressions via electric
conductance are not considered. Specially, since the gauge field
obtains a time-dependent mass, as is evident in Eq. (\ref{Ak-Eq}),
the effects from parametric resonance and significant amplification
of magnetic field during preheating stage may play some important
roles \cite{Davis:2000zp, Bassett:2000aw, Giovannini:2000wta}. This
may help to relax the bounds above by few orders of magnitude.

In this work we have turned off the background gauge field classically. Although this is a consistent solution, but it is not the most natural solution and requires an additional fine-tuning in our model. In a future work \cite {Emami} we study the case in our model where the background gauge field is not turned off classically. For this, one has to search for the parameter space where the gauge-field energy density does not dominate over the inflaton field energy density such that the produced anisotropies are within the observational bounds from CMB. This is similar to the analysis of \cite{Watanabe:2009ct} except that in their model there is no gauge coupling, $e=0$, and $\beta=0$.

\section{Acknowledgement}

We would like to thank Bruce Bassett, Robert Brandenberger, Jim
Cline, Paolo Creminelli,  Massimo Giovannini, Lev Kofman
and Shahin Sheikh-Jabbari for valuable discussions and comments.
We also thank the anonymous PRD referee for the insightful critical comments and for useful suggestions on the draft. H.F. would like
to thank Perimeter Institute and McGill University for hospitality
where this work was completed.



\section*{References}
\begin{thebibliography}{}

\bibitem{Grasso:2000wj}
  D.~Grasso and H.~R.~Rubinstein,
  ``Magnetic fields in the early universe,''
  Phys.\ Rept.\  {\bf 348}, 163 (2001)
  [arXiv:astro-ph/0009061].

\bibitem{Widrow:2002ud}
  L.~M.~Widrow,
  ``Origin of Galactic and Extragalactic Magnetic Fields,''
  Rev.\ Mod.\ Phys.\  {\bf 74}, 775 (2002)
  [arXiv:astro-ph/0207240].

\bibitem{Giovannini:2002sv}
  M.~Giovannini,
  ``Primordial magnetic fields,''
  arXiv:hep-ph/0208152;\\
  M.~Giovannini,
  ``Magnetic fields, strings and cosmology,''
  Lect.\ Notes Phys.\  {\bf 737}, 863 (2008)
  [arXiv:astro-ph/0612378].

\bibitem{Davis:1999bt}
  A.~C.~Davis, M.~Lilley and O.~Tornkvist,
  ``Relaxing the Bounds on Primordial Magnetic Seed Fields,''
  Phys.\ Rev.\  D {\bf 60}, 021301 (1999)
  [arXiv:astro-ph/9904022].

\bibitem{Turner:1987bw}
  M.~S.~Turner and L.~M.~Widrow,
  ``Inflation Produced, Large Scale Magnetic Fields,''
  Phys.\ Rev.\  D {\bf 37}, 2743 (1988).

\bibitem{Ratra:1991bn}
  B.~Ratra,
  ``Cosmological 'seed' magnetic field from inflation,''
  Astrophys.\ J.\  {\bf 391}, L1 (1992).


\bibitem{Dolgov:1993vg}
  A.~Dolgov,
  ``Breaking Of Conformal Invariance And Electromagnetic Field Generation In
  The Universe,''
  Phys.\ Rev.\  D {\bf 48}, 2499 (1993)
  [arXiv:hep-ph/9301280].

\bibitem{Lemoine:1995vj}
  D.~Lemoine and M.~Lemoine,
  ``Primordial magnetic fields in string cosmology,''
  Phys.\ Rev.\  D {\bf 52}, 1955 (1995).

\bibitem{Gasperini:1995dh}
  M.~Gasperini, M.~Giovannini and G.~Veneziano,
  ``Primordial magnetic fields from string cosmology,''
  Phys.\ Rev.\ Lett.\  {\bf 75}, 3796 (1995)
  [arXiv:hep-th/9504083].

\bibitem{Calzetta:1997ku}
  E.~A.~Calzetta, A.~Kandus and F.~D.~Mazzitelli,
  ``Primordial magnetic fields induced by cosmological particle creation,''
  Phys.\ Rev.\  D {\bf 57}, 7139 (1998)
  [arXiv:astro-ph/9707220];\\
  E.~A.~Calzetta and A.~Kandus,
  ``Self consistent estimates of magnetic fields from reheating,''
  Phys.\ Rev.\  D {\bf 65}, 063004 (2002)
  [arXiv:astro-ph/0110341].

\bibitem{Finelli:2000sh}
  F.~Finelli and A.~Gruppuso,
  ``Resonant amplification of gauge fields in expanding universe,''
  Phys.\ Lett.\  B {\bf 502}, 216 (2001)
  [arXiv:hep-ph/0001231].

\bibitem{Giovannini:2000dj}
  M.~Giovannini and M.~E.~Shaposhnikov,
  ``Primordial magnetic fields from inflation?,''
  Phys.\ Rev.\  D {\bf 62}, 103512 (2000)
  [arXiv:hep-ph/0004269].

\bibitem{Giovannini:2000wta}
  M.~Giovannini and M.~E.~Shaposhnikov,
  ``Primordial magnetic fields from inflation?,''
  arXiv:hep-ph/0011105.

\bibitem{Davis:2000zp}
  A.~C.~Davis, K.~Dimopoulos, T.~Prokopec and O.~Tornkvist,
  ``Primordial spectrum of gauge fields from inflation,''
  Phys.\ Lett.\  B {\bf 501}, 165 (2001)
  [Phys.\ Rev.\ Focus {\bf 10}, STORY9 (2002)]
  [arXiv:astro-ph/0007214].

\bibitem{Gasperini:2000tw}
  M.~Gasperini,
  ``A new mechanism for the generation of primordial seeds for the cosmic
  magnetic fields,''
  Phys.\ Rev.\  D {\bf 63}, 047301 (2001)
  [arXiv:astro-ph/0009476].

\bibitem{Bassett:2000aw}
  B.~A.~Bassett, G.~Pollifrone, S.~Tsujikawa and F.~Viniegra,
  ``Preheating as cosmic magnetic dynamo,''
  Phys.\ Rev.\  D {\bf 63}, 103515 (2001)
  [arXiv:astro-ph/0010628].

\bibitem{Ashoorioon:2004rs}
  A.~Ashoorioon and R.~B.~Mann,
  ``Generation of cosmological seed magnetic fields from inflation with
  cutoff,''
  Phys.\ Rev.\  D {\bf 71}, 103509 (2005)
  [arXiv:gr-qc/0410053].


\bibitem{Giovannini:2001nh}
  M.~Giovannini,
  ``On the variation of the gauge couplings during inflation,''
  Phys.\ Rev.\  D {\bf 64}, 061301 (2001)
  [arXiv:astro-ph/0104290].

\bibitem{Bamba:2003av}
  K.~Bamba and J.~Yokoyama,
  ``Large-scale magnetic fields from inflation in dilaton electromagnetism,''
  Phys.\ Rev.\  D {\bf 69}, 043507 (2004)
  [arXiv:astro-ph/0310824].

\bibitem{Bamba:2008hr}
  K.~Bamba, C.~Q.~Geng and S.~H.~Ho,
  ``Large-scale magnetic fields from inflation due to Chern-Simons-like
  effective interaction,''
  JCAP {\bf 0811}, 013 (2008)
  [arXiv:0806.1856 [astro-ph]];\\
  K.~Bamba, N.~Ohta and S.~Tsujikawa,
  ``Generic estimates for magnetic fields generated during inflation including
  Dirac-Born-Infeld theories,''
  Phys.\ Rev.\  D {\bf 78}, 043524 (2008)
  [arXiv:0805.3862 [astro-ph]];\\
  K.~Bamba,
  ``The interrelation between the generation of large-scale electric fields and
  that of large-scale magnetic fields during inflation,''
  JCAP {\bf 0710}, 015 (2007)
  [arXiv:0710.1906 [astro-ph]];\\
  K.~Bamba,
  ``Property of the spectrum of large-scale magnetic fields from inflation,''
  Phys.\ Rev.\  D {\bf 75}, 083516 (2007)
  [arXiv:astro-ph/0703647].

\bibitem{Ganjali:2005sr}
  M.~A.~Ganjali,
  ``DBI with primordial magnetic field in the sky,''
  JHEP {\bf 0509}, 004 (2005)
  [arXiv:hep-th/0509032].

\bibitem{Bamba:2006ga}
  K.~Bamba and M.~Sasaki,
  ``Large-scale magnetic fields in the inflationary universe,''
  JCAP {\bf 0702}, 030 (2007)
  [arXiv:astro-ph/0611701].

\bibitem{Giovannini:2007rh}
  M.~Giovannini,
  ``Magnetogenesis, spectator fields and CMB signatures,''
  Phys.\ Lett.\  B {\bf 659}, 661 (2008)
  [arXiv:0711.3273 [astro-ph]].

  \bibitem{BY}
  K.~Bamba and J.~Yokoyama,
 ``Large-scale magnetic fields from dilaton inflation in noncommutative
 spacetime,'' Phys.\ Rev.\  D {\bf 70} (2004) 083508.
 \bibitem{BRM}
O.~Bertolami and R.~Monteiro,
 ``Varying electromagnetic coupling and primordial magnetic fields,''
 Phys.\ Rev.\  D {\bf 71} (2005) 123525.
 \bibitem{Salim}
J.~M.~Salim, N.~Souza, S.~E.~Perez Bergliaffa and T.~Prokopec,
 ``Creation of cosmological magnetic fields in a bouncing cosmology,''
 JCAP {\bf 0704} (2007) 011.
 \bibitem{Martin}
J.~Martin and J.~Yokoyama,
 ``Generation of Large-Scale Magnetic Fields in Single-Field
 Inflation,'' JCAP {\bf 0801} (2008) 025.
 \bibitem{Bamba-Odnitsov}
K.~Bamba and S.~D.~Odintsov,
 ``Inflation and late-time cosmic acceleration in non-minimal
 Maxwell-$F(R)$ gravity and the generation of large-scale magnetic
 fields,'' JCAP {\bf 0804} (2008) 024.
 \bibitem{Bamba-G-H}
K.~Bamba, C.~Q.~Geng and S.~H.~Ho,
 ``Large-scale magnetic fields from inflation due to Chern-Simons-like
 effective interaction,'' JCAP {\bf 0811} (2008) 013.

  \bibitem{Garretson}
W.~D.~Garretson, G.~B.~Field and S.~M.~Carroll,
 ``Primordial magnetic fields from pseudoGoldstone bosons,'' Phys.\
 Rev.\  D {\bf 46} (1992) 5346.
 \bibitem{MS}
F.~D.~Mazzitelli and F.~M.~Spedalieri,
 ``Scalar electrodynamics and primordial magnetic fields,'' Phys.\
 Rev.\  D {\bf 52} (1995) 6694.
 \bibitem{Novello}
M.~Novello, L.~A.~R.~Oliveira and J.~M.~Salim,
 ``Direct Electrogravitational Couplings And The Behavior Of Primordial
 Large Scale Magnetic Fields,'' Class.\ Quant.\ Grav.\  {\bf 13}
 (1996) 1089. 
  \bibitem{Mota}
O.~Bertolami and D.~F.~Mota,
 ``Primordial magnetic fields via spontaneous breaking of Lorentz invariance,'' Phys.\ Lett.\  B {\bf 455} (1999) 96.
  \bibitem{Massimmo}
M.~Giovannini,
 ``Magnetogenesis and the dynamics of internal dimensions,'' Phys.\
 Rev.\  D {\bf 62} (2000) 123505.
  \bibitem{Kostas}
K.~Dimopoulos, T.~Prokopec, O.~Tornkvist and A.~C.~Davis,
 ``Natural magnetogenesis from inflation,'' Phys.\ Rev.\  D {\bf 65}
 (2002) 063505.
  \bibitem{Prokopec}
T.~Prokopec and E.~Puchwein,
 ``Photon mass generation during inflation: de Sitter invariant case,''
 JCAP {\bf 0404} (2004) 007.
T.~Prokopec and E.~Puchwein, ``Nearly minimal magnetogenesis,'' Phys.\ Rev.\  D {\bf 70} (2004)  043004.
  \bibitem{Anupam}
K.~Enqvist, A.~Jokinen and A.~Mazumdar,
 ``Seed perturbations for primordial magnetic fields from MSSM flat
 directions,'' JCAP {\bf 0411} (2004) 001.
  \bibitem{Garousi}
M.~R.~Garousi, M.~Sami and S.~Tsujikawa,
 ``Generation of electromagnetic fields in string cosmology with a
 massive scalar field on the anti D-brane,'' Phys.\ Lett.\  B {\bf 606}
 (2005) 1.
  \bibitem{Mardiz}
J.~E.~Madriz Aguilar and M.~Bellini,
 ``Stochastic gravitoelectromagnetic inflation,'' Phys.\ Lett.\  B
 {\bf 642} (2006) 302.
  \bibitem{Bellini}
F.~Agustin Membiela and M.~Bellini,
 ``Power spectrum of large-scale magnetic fields from
 Gravitoelectromagnetic inflation with a decaying cosmological
 parameter,'' arXiv: 0712.3032 [hep-th].
  \bibitem{Cea}
L.~Campanelli, P.~Cea, G.~L.~Fogli and L.~Tedesco,
 ``Inflation-Produced Magnetic Fields in Nonlinear Electrodynamics,''
 Phys.\ Rev.\  D {\bf 77} (2008) 043001;
L.~Campanelli, P.~Cea, G.~L.~Fogli and L.~Tedesco,
 ``Inflation-Produced Magnetic Fields in $R^n F^2$ and $I F^2$ models,''
 Phys.\ Rev.\  D {\bf 77} (2008) 123002.

\bibitem{Campanelli:2008kh}
  L.~Campanelli,
  ``Helical Magnetic Fields from Inflation,''
  Int.\ J.\ Mod.\ Phys.\  D {\bf 18}, 1395 (2009)
  [arXiv:0805.0575 [astro-ph]].
\bibitem{Demozzi:2009fu}
  V.~Demozzi, V.~Mukhanov and H.~Rubinstein,
  ``Magnetic fields from inflation?,''
  arXiv:0907.1030 [astro-ph.CO].


\bibitem{Kanno:2009ei}
  S.~Kanno, J.~Soda and M.~a.~Watanabe,
  ``Cosmological Magnetic Fields from Inflation and Backreaction,''
  arXiv:0908.3509 [astro-ph.CO].

\bibitem{Dolgov:1993mu}
  A.~Dolgov and J.~Silk,
  ``Electric charge asymmetry of the universe and magnetic field generation,''
  Phys.\ Rev.\  D {\bf 47}, 3144 (1993).
  
\bibitem{Emami}
R. Emami, H. Firouzjahi, S. Movahed and M. Zarei, work in progress .

\bibitem{Watanabe:2009ct}
  M.~a.~Watanabe, S.~Kanno and J.~Soda,
  ``Inflationary Universe with Anisotropic Hair,''
  Phys.\ Rev.\ Lett.\  {\bf 102}, 191302 (2009)
  [arXiv:0902.2833 [hep-th]].



\end{thebibliography}
\end{document}